\documentclass[12pt]{article}

\usepackage{amsmath,amssymb}
\usepackage{amsfonts}
\usepackage{graphicx}
\usepackage{a4wide}

\usepackage[T2A]{fontenc}
\usepackage[cp866]{inputenc}
\usepackage[russian]{babel}

\begin{document}

\noindent {\it Problems of Information Transmission},\\
\noindent vol. 46, no. 2, pp. 3--23, 2010.

\begin{center} {\bf M. V. Burnashev\footnote[1]{The research
described in this publication
was made possible in part by the Russian Fund for Fundamental
Research (project numbers 06-01-00226 and 09-01-00536).}, H.
Yamamoto} \end{center}

\vskip 0.4cm

\begin{center}
{\large\bf ON RELIABILITY FUNCTION OF BSC \\ WITH NOISY FEEDBACK}
\end{center}

{\begin{quotation} \normalsize For information transmission a
binary symmetric channel is used. There is also another noisy
binary symmetric channel (feedback channel), and the transmitter
observes without delay all the outputs of the forward channel via
that feedback channel. The transmission of an exponential number
of messages (i.e. the transmission rate is positive) is considered.
The achievable decoding error exponent for such a combination
of channels is investigated. It is shown that if the crossover
probability of the feedback channel is less than a certain
positive value, then the achievable error exponent is better than
the decoding error exponent of the channel without feedback.
\end{quotation}}

\vskip 0.7cm

\begin{center}
{\bf \S\,1. Introduction and main results}
\end{center}

The binary symmetric channel ${\rm BSC}(p)$ with crossover
probability $0 < p < 1/2$ (and $q = 1-p$) is considered. It is
assumed that there is also the feedback ${\rm BSC}(p_{1})$ channel,
and the transmitter observes (without delay) all outputs of the
forward ${\rm BSC}(p)$ channel via that noisy feedback channel.
No coding is used in the feedback channel (i.e. the receiver simply
resends to the transmitter all received outputs). In other words,
the feedback channel is ``passive'' (see Fig. 1).

\begin{picture}(50,70)(30,-20)
\put(80,5){\framebox(70,20){Transm.}}
\put(190,5){\framebox(100,20){BSC$(p)$}}
\put(190,-35){\framebox(100,20){BSC$(p_{1})$}}
\put(350,5){\framebox(70,20){Receiver}}

\put(290,15){\vector(1,0){60}}
\put(150,15){\vector(1,0){40}}
\put(60,-25){\line(1,0){130}}
\put(30,15){\vector(1,0){50}}
\put(320,-25){\vector(-1,0){30}}
\put(320,-25){\line(0,1){40}}
\put(60,-25){\vector(0,1){40}}
\put(420,15){\vector(1,0){30}}

\put(170,20){$x$}
\put(310,20){$y$}
\put(170,-20){$x'$}
\end{picture}

\vskip 1.0cm

\begin{center}
{Fig. \,1. Channel model}
\end{center}

We consider the case when the overall transmission time $n$ and
$M= e^{Rn}$ equiprobable messages $\{\theta_{1},\ldots,\theta_{M}\}$
are given. After the moment $n$, the receiver makes a decision
${\hat \theta}$ on the message transmitted. We are interested in the
best possible decoding error exponent (and whether it can exceed the
similar exponent  of the channel without feedback).

Such model was considered in \cite{BY1}, where the case of a
nonexponential (on $n$) number $M$ (i.e. $R = 0$) was investigated.
In the paper we consider the case $M = e^{Rn},\,R > 0$,
strengthening methods of \cite{BY1}. The main difference is that
since now $M$ is exponential in $n$, we will need much more
accurate investigation of the decoding error probability.
Moreover, if $M$ is nonexponential in $n$, then we know the best
code for use during phase I - it is an ``almost equidistant'' code
(i.e. all its codeword distances equal $n/2+o(n)$). If $R > 0$ then
we do not know such best code, and for that reason we choose that
code randomly.

Some results for channels with noiseless feedback can be
found in [2--12], and in the noisy feedback case -- in
\cite{DrapSah1, KimLapW} (see also discussion in \cite{BY1}).

We show that if the crossover probability $p_{1}$ of the feedback
channel ${\rm BSC}(p_{1})$ is less then the certain positive value
$p_{0}(p,R)$, then it is possible to improve the best error
exponent $E(R,p)$ of ${\rm BSC}(p)$ without feedback. The
transmission method with one ``switching'' moment, giving such an
improvement, is described in  \S\,4. It is similar to the method
used in \cite{BY1}.

We will need some definitions and notations.
For $L=1,2,\ldots$ define the critical rates
$R_{{\rm crit}, 1}(p) > R_{{\rm crit}, 2}(p)> \ldots$
\cite{E, G2, G1}
\begin{equation}\label{Rcrit}
R_{{\rm crit}, L}(p) = \ln 2 - h\left[\frac{p^{1/(L+1)}}
{p^{1/(L+1)} + q^{1/(L+1)}}\right],
\end{equation}
where $h(x) = -x\ln x - (1-x)\ln (1-x)$.
For $L = 1$ we omit the index $L$ and simply write
$R_{\rm crit}(p) = R_{{\rm crit}, 1}(p),\,E(R,p) = E(R,p,1)$, etc.

Define the new critical rate $R_{2} = R_{2}(p)$
as the unique root of the equation \cite{Bur3}
$$
\min_{\substack
{0 \leq \tau \leq \alpha \leq 1/2 \\
h(\alpha) - h(\tau) = \ln 2-R_{2}}} \frac{\alpha(1-\alpha) -
\tau(1-\tau)} {1+2\sqrt{\tau(1-\tau)}} = \frac{\sqrt{pq}}{1 +
2\sqrt{pq}}\,.
$$
Then $0 < R_{2}(p) < R_{\rm crit}(p),\,0 < p < 1/2$.

Denote by $C(p) = \ln 2 -h(p)$ the capacity of the ${\rm BSC}(p)$,
and by $E_{\rm sp}(R,p)$ the sphere-packing exponent
$$
\begin{gathered}
E_{\rm sp}(R,p) = D\left(\delta_{GV}(R)\,\| \,p\right), \\
D(x\,\| \, y) = x\ln \frac{x}{y} + (1-x)\ln \frac{1-x}{1-y},
\end{gathered}
$$
where $\delta_{GV}(R) \leq 1/2$ is defined by the relation
$$
\ln 2 - R = h(\delta_{GV}(R)).
$$

Denote by $E(R,p)$ the best decoding error exponent (the reliability
function) of ${\rm BSC}(p)$ without feedback. For
$R_{2}(p) \leq R \leq C(p)$, and $R = 0$ the function $E(R,p)$
is known exactly \cite{E, Bur3}:
\begin{equation}\label{main2}
\begin{gathered}
E(R,p) = E_{\rm r}(R,p) = \left\{
\begin{array}{ll}
\ln 2 - \ln\left(1+2\sqrt{pq}\right) - R, &
R_{2}(p) \leq R \leq R_{\rm crit}(p), \\
E_{\rm sp}(R,p), &  R_{\rm crit}(p) \leq R \leq C(p),
\end{array}
\right. \\
E(0,p) = E_{\rm ex}(0,p) = \frac{1}{4}\ln \frac{1}{4pq},
\end{gathered}
\end{equation}
where $E_{\rm r}(R,p), E_{\rm ex}(R,p)$ -- ``random coding'' bounds
\cite{E,G2, G1} (see \S\,6).

For $0 < R < R_{2}(p)$ there are known only lower and upper bounds
for the function $E(R,p)$. To describe the best known lower bound
(the exponent $E_{\rm ex}(R,p)$ of random coding with
``expurgation''), introduce the rate $R_{\rm min}(p)$
(see (\ref{Rmin12})). Then
$0 < R_{\rm min}(p) < R_{2}(p) < R_{\rm crit}(p),\, 0 < p < 1/2$,
and the best known lower bound \cite{G2,G1} has the form
\begin{equation}\label{Elow1}
E(R,p) \geq E_{\rm ex}(R,p) = \left\{
\begin{array}{ll}
-\delta_{GV}(R)\ln\sqrt{4pq}, &  0 < R \leq R_{\rm min}(p), \\
\ln 2 - \ln\left(1+2\sqrt{pq}\right) - R, &
R_{\rm min}(p) \leq R < R_{2}(p).
\end{array}
\right.
\end{equation}

Denote by $E(R,p,L)$ the best list size $L$ decoding error exponent
of ${\rm BSC}(p)$ without feedback. It is known that
$E(R,p,L) = E_{\rm r}(R,p,L) = E_{\rm sp}(R,p), \,
R_{{\rm crit}, L}(p) \leq R < C(p)$ \cite{E,G2, G1} and
$E(0,p,L) = E_{\rm ex}(0,p,L)$ \cite{Blin2}, where the
``random coding'' $E_{\rm r}(R,p,L)$ and the ``random coding with
expurgation'' $E_{\rm ex}(R,p,L)$ bounds are described in \S\,6.

For $0 < R < R_{{\rm crit}, L}(p)$ the best known lower bound for
$E(R,p,L)$ has the form \cite{G2,G1}
\begin{equation}\label{Elow2}
\begin{gathered}
E(R,p,L) \geq E_{\rm ex}(R,p,L), \quad 0 < R < R_{{\rm crit}, L}(p).
\end{gathered}
\end{equation}
We also have $E_{\rm ex}(R,p,L) = E_{\rm r}(R,p,L),\,
R_{{\rm min}, L}(p) \leq R \leq R_{{\rm crit}, L}(p)$ (see
(\ref{RminL})). Denote
\begin{equation}\label{Elow}
\begin{gathered}
E_{\rm low}(R,p,L) = \max\{E_{\rm r}(R,p,L),
E_{\rm ex}(R,p,L)\}.
\end{gathered}
\end{equation}

Denote by $F(R,p)$ the best decoding error exponent of
${\rm BSC}(p)$ with noiseless feedback. Then
$$
\begin{array}{lll}
 E(R,p) = F(R,p) = E_{\rm sp}(R,p), & R_{\rm crit}(p) \leq R \leq
C(p) & \cite{Dob},  \\
E(R,p) \leq F(R,p) \leq E_{\rm sp}(R,p), & 0 < R < R_{\rm crit}(p) &
\cite{Dob}, \\
E(0,p) < F(0,p) = -\ln\left(p^{1/3}q^{2/3} + p^{2/3}q^{1/3}\right) &
& \cite{Ber1}.
\end{array}
$$

Denote by $F(R,p,p_{1})$ the best decoding error exponent of
${\rm BSC}(p)$ with the noisy ${\rm BSC}(p_{1})$ feedback channel.
Clearly, $E(R,p) \leq F(R,p,p_{1}) \leq F(R,p)$ for all $p,p_{1}$.
In particular, $F(R,p,0) = F(R,p),\, F(R,p,1/2) = E(R,p)$.

Denote by $E_{2}(p)$ the best error exponent for two codewords over
${\rm BSC}(p)$ (clearly, it remains the same for the channel with
noiseless feedback as well)
\begin{equation}\label{defE2}
\begin{gathered}
E_{2}(p) = \frac{1}{2}\ln \frac{1}{4pq}.
\end{gathered}
\end{equation}

Denote by $F_{1}(R,p,p_{1})$ the decoding error exponent of the
transmission method described in \S\,4 (with one switching moment).
The inequality $F_{1}(R,p,p_{1}) > E(R,p)$ is possible only when
$R< R_{\rm crit}(p)$.

To describe the function $p_{0}(R,p)$ of the critical noise level
in the feedback channel, introduce the function
\begin{equation}\label{deft0}
\begin{gathered}
t_{0}(R,p) = \frac{3\left[E_{\rm low}(R,p,2)- E_{\rm low}(R,p)
\right]}{\ln(q/p)},
\end{gathered}
\end{equation}
where $E_{\rm low}(R,p,2),\,E_{\rm low}(R,p) = E_{\rm low}(R,p,1)$
are defined in (\ref{Elow}).

The function $t_{0}(R,p)$ monotonically decreases on $R$. For a
given $R \geq 0$ it first increases on $p$, and then decreases.
Moreover,
$$
\max_{R,p}t_{0}(R,p) = \max_{p}t_{0}(0,p) \approx t_{0}(0,0.0124)
\approx 0.1322.
$$

Introduce the function $p_{0} = p_{0}(R,p) \leq t_{0}(R,p)$ as the
unique root of the equation
\begin{equation}\label{defp_{1}}
D\left(t_{0}(R,p)\,\| \,p_{0}\right) = 2R.
\end{equation}
In particular,
$$
p_{0}(0,p) = t_{0}(0,p) = \frac{3\left[\ln 4 -
3\ln\left(p^{1/3} + q^{1/3}\right)\right]}{4\ln(q/p)}.
$$
Define also $t_{1} = t_{1}(R,p_{1}) \geq p_{1}$ as the unique root
of the equation
\begin{equation}\label{deft_{1}}
D\left(t_{1}\,\| \,p_{1}\right) = 2R.
\end{equation}

The main result of the paper represents

{T h e o r e m \,1}. {\it If $R < R_{\rm crit}(p)$ and
$p_{1} < p_{0}(R,p)$, then
\begin{equation}\label{Theor1}
\begin{gathered}
F_{1}(R,p,p_{1}) \geq \max_{0 \leq \gamma \leq 1}
T(R,p,p_{1},\gamma) > \left\{
\begin{array}{ll}
  E_{\rm ex}(R,p),  & 0 \leq R \leq R_{2}(p), \\
  E(R,p), & R_{2}(p) \leq R < R_{\rm crit}(p),
\end{array}
\right.
\end{gathered}
\end{equation}
where}
\begin{equation}\label{Theor1a}
\begin{gathered}
T = \min\bigg\{\gamma E_{\rm low}(R/\gamma,p,2) -
\frac{\gamma t_{1}(R/\gamma,p_{1})}{3}\ln\frac{q}{p},\,
\gamma E_{\rm low}(R/\gamma,p) + (1-\gamma)E_{2}(p)\bigg\}.
\end{gathered}
\end{equation}

In other words, for any $R < R_{\rm crit}(p)$ and
$p_{1} < p_{0}(R,p)$ the function $F_{1}(R,p,p_{1})$ is bigger
(i.e. better) than the best known lower bound for the decoding error
exponent of ${\rm BSC}(p)$ without feedback.

Moreover, there exists the positive function $p_{2}(R,p)$ such that
the following result holds.

C o r o l l a r y \,1. {\it If $R < R_{\rm crit}(p)$ and
$p_{1} < p_{2}(R,p)$, then}
\begin{equation}\label{Theor1b}
\begin{gathered}
F_{1}(R,p,p_{1}) \geq \max_{0 \leq \gamma \leq 1}
T(R,p,p_{1},\gamma) > E(R,p).
\end{gathered}
\end{equation}

This result follows from the proof of the Theorem 2 (see \S 3) and
the fact that the function $T(R,p,p_{1},\gamma)$ is continuous on
$p_{1}$.

{\it Remark} 1. We do not try to find the best function $p_{0}(R,p)$,
limiting ourselves to rather simple estimates for it.

On Fig. 2. the plot of the function $p_{0}(R,p)$ for $p = 0.01$ is
given ($R_{\rm crit} \approx 0.387$). Note that here
$p_{0}(R,p) > p$ for small $R$.

It is more convenient for us to investigate first the function
$F_{1}(R,p,p_{1})$ for $p_{1} = 0$, i.e. for the channel with
noiseless feedback. Then the next result holds.

{T h e o r e m \,2}. {\it If $0 < p < 1/2,\,R < R_{\rm crit}(p)$,
then
\begin{equation}\label{Theor2}
F_{1}(R,p,0) = F_{1}(R,p) \geq \gamma_{0}
E_{\rm low}(R/\gamma_{0},p,2) > E(R,p),
\end{equation}
where $\gamma_{0} \in (R/R_{\rm crit}(p),1)$ is the largest root of
the equation} (\ref{Pe3}).

{\it Remark} 2. If $p_{1} \to 0$, then the relations (\ref{Theor1}),
(\ref{Theor1a}) turn into the similar relation (\ref{Theor2}) for
the channel with noiseless feedback (see also remark 6 in \S 4).

{\it Remark} 3. The transmission method described in \S\,4, reduces
the problem to testing of two most probable (at a fixed moment)
messages. Such strategy is not optimal even for one switching
moment (at least, if $p_{1}$ is very small). But it is relatively
simple for investigation, and it gives already a reasonable
improvement over the channel without feedback.

{\it Remark} 4. In the preliminary publication
\cite[Proposition]{BurYam1} it was claimed that $p_{0}(R,p) = 1/2$
for some range of rates $R$. In the proof of that result a
miscalculation was found.

Below in \S\,2 informal description of the transmission method is
given. In \S\,3 the \\ transmission method with one switching moment
in the case of the channel with noiseless feedback is described and
analyzed and the Theorem 2 is proved. In \S\,4 that method (slightly
modified) is investigated for the channel with noisy feedback and
the Theorem 1 is proved. In \S\,5 it is clarified for which $p_{1}$
noisy feedback behaves approximately like noiseless. A part of
formulas used and some auxiliary results are presented in \S\,6.

A preliminary (and simplified) paper variant (without detailed
proofs) was published in \cite{BurYam1}.

\begin{center}
{\bf \S\,2. Informal description of the transmission method}
\end{center}

We use the transmission method with one fixed switching moment at
which the coding function is changed. That method is based on
one idea and one useful observation.

\underline{Idea}. It is based on the inequality which follows from
(\ref{idea2})
\begin{equation}\label{idea1}
E_{\rm ex}(R,p) < E_{\rm low}(R,p,2), \qquad R < R_{\rm crit}(p).
\end{equation}
Considering only $R < R_{\rm crit}(p)$ we choose some positive
$\gamma < 1$ and partition the total transmission period $[1,n]$ on
two phases: $[1,\gamma n]$ (phase I) and $(\gamma n, n]$ (phase II)
(at first we may think that $\gamma$ is rather close to one).

On phase I (i.e. on $[0,\gamma n]$) we use the ``best'' code of
$M$ codewords $\{\mbox{\boldmath $x$}_{i}\}$ of length $\gamma n$
(see below). On that phase the transmitter only observes (via the
feedback channel) outputs of the forward channel, but does not
change the coding function. We set the value
$\gamma = \gamma(R,p)$ such that
\begin{equation}\label{idea1a}
E_{\rm ex}(R,p) < \gamma E_{\rm low}(R/\gamma,p,2)\,, \qquad
R < R_{\rm crit}(p)
\end{equation}
(it is always possible due to continuity of the function
$\gamma E_{\rm low}(R/\gamma,p,2)$ on $\gamma$ and the condition
(\ref{idea1})). After phase I (at moment $\gamma n$) the receiver
selects two most probable messages $\theta_{i},\theta_{j}$. By the
condition (\ref{idea1a}), the exponent of the probability that
the true message $\theta_{\rm true}$ is not among the chosen
messages $\theta_{i},\theta_{j}$, will be larger (i.e. better) than
$E_{\rm ex}(R,p)$. Assume that by some means the transmitter is also
able to recover those two most probable messages
$\theta_{i},\theta_{j}$ (it is certainly so in the noiseless
feedback case). Then, on phase II (i.e. on $(\gamma n, n]$) the
transmitter only helps the receiver to decide between those two most
probable messages $\theta_{i},\theta_{j}$, using two opposite
codewords of length $(1-\gamma) n$. The error exponent $E_{2}(p)$
(see (\ref{defE2})) on that phase is better than all other exponents
involved. As a result, it gives the overall decoding error exponent
better than $E_{\rm ex}(R,p)$.

It remains us to find the way the transmitter will able to recover
those two most probable messages $\theta_{i},\theta_{j}$. It may
seem that it is always possible if the value $p_{1}$ is sufficiently
small. But it is not true. With high probability (even close to one)
the second $\theta_{j}$ and the third $\theta_{k}$ most probable
messages will be approximately equiprobable, and then, for any
$p_{1} > 0$, the transmitter will not be able to rank them
correctly (due to noise in the feedback channel).

\underline{Observation}. Fortunately, in that case (with high
probability)
the most probable \\ message $\theta_{i}$ will be much more probable
than the second most probable message $\theta_{j}$. In such case
the receiver makes a decision immediately after phase I (in favor of
the most probable message $\theta_{i}$), and it ignores all next
signals from the transmitter.

The description given is rather intuitive, and it should be
checked analytically (which is done below).


\begin{center}
{\bf \S\,3. Channel with noiseless feedback. Proof of Theorem 2}
\end{center}

For simplicity, we start with the noiseless feedback case and
describe formally the \\ transmission method which (after some
modification) will be used for noisy feedback as well. Moreover, in
the noisy feedback case we will need some formulas from the
noiseless feedback case.

Denote by $F_{1}(R,p) = F_{1}(R,p,0)$ the decoding error exponent of
the transmission method described below (with one switching moment).

P r o o f \;o f \;T h e o r e m \,2. We consider $M = e^{Rn}$
messages $\theta_{1},\ldots,\theta_{M}$. Using some
$\gamma \in [0,1]$ (it will be chosen later), we partition the total
transmission time $[1,n]$ on two phases: $[1,\gamma n]$ (phase I)
and $(\gamma n, n]$ (phase II). We perform as follows.

1) On phase I (i.e. on $[1,\gamma n]$) we use the ``best'' code of
$M$ codewords $\{\mbox{\boldmath $x$}_{i}\}$ of length $\gamma n$
(see below). On that phase the transmitter only observes (via the
feedback channel) outputs of the forward channel, but does not
change the coding function.

2) Let $\mbox{\boldmath $x$}$ be the transmitted codeword (of length
$\gamma n$) and  $\mbox{\boldmath $y$}$ be the received (by the
receiver) block. After phase I, based on the block
$\mbox{\boldmath $y$}$, the transmitter selects two messages
$\theta_{i},\theta_{j}$ (codewords
$\mbox{\boldmath $x$}_{i},\mbox{\boldmath $x$}_{j}$) which are the
most probable for the receiver, and ignores all the remaining
messages $\{\theta_{k}\}$. If among the selected messages
$\theta_{i},\theta_{j}$ there is the true message
$\theta_{\rm true}$, then on phase II (i.e. on $(\gamma n, n]$) the
transmitter only helps the receiver to decide between those two most
probable messages $\theta_{i},\theta_{j}$, using two opposite
codewords of length $(1-\gamma) n$. If the true  message
$\theta_{\rm true}$ is not among those two selected messages, then
the transmitter sends an arbitrary block. After moment $n$ the
receiver makes a decision in favor of the most probable of those two
remaining messages $\theta_{i},\theta_{j}$ (based on all received
on $[1,n]$ signals).

Clearly, a decoding error occurs in the following two cases.

1) After phase I the true message is not among two most probable
messages. We denote that probability by $P_{1}$.

2) After phase I the true message is among two most probable, but
after phase II it is not the most probable. We denote that
probability by $P_{20}$.

Then for the total decoding error probability $P_{\rm e}$ we have
\begin{equation}\label{Petotal}
P_{\rm e} \leq P_{1} + P_{20}\,.
\end{equation}

On phase I (of length $\gamma n$) we use a code having small two
decoding error probabilities: usual and when decoding with list size
$L = 2$. Then there exists a code such that for $P_{1}$ we have
(see \S\,6)
\begin{equation}\label{Pe1list}
\frac{1}{n}\ln \frac{1}{P_{1}} \geq \gamma E_{\rm low}(R/\gamma,p,2)
+ o(1), \qquad n \to \infty\,.
\end{equation}

Now we evaluate the probability $P_{20}$. Denote by
$d\left(\mbox{\boldmath $x$},\mbox{\boldmath $y$}\right)$ the
Hamming distance between $\mbox{\boldmath $x$}$ and
$\mbox{\boldmath $y$}$, and
$d_{ij} = d\left(\mbox{\boldmath $x$}_{i},\mbox{\boldmath $x$}_{j}
\right)$. On phase I (of length $\gamma n$) the distances among
codewords are $\{d_{ij}\}$. On phase II (of length $(1-\gamma) n$)
the distance between two remaining codewords equals $(1-\gamma)n$.
Therefore, the  total distance between the true and the concurrent
codewords equals $d_{ij} + (1-\gamma)n$. Then there exists a code
such that (see derivation in \S\,6)
\begin{equation}\label{Pe2}
\begin{gathered}
\frac{1}{n}\ln\frac{1}{P_{20}} \geq \gamma E_{\rm low}(R/\gamma,p) +
(1-\gamma)E_{2}(p) + o(1).
\end{gathered}
\end{equation}
Moreover, there exists a code for which both relations
(\ref{Pe1list}) and (\ref{Pe2}) are fulfilled (see \S\,6).
Then from (\ref{Petotal})--(\ref{Pe2}) we have
$$
\begin{gathered}
\frac{1}{n}\ln \frac{1}{P_{\rm e}} \geq \frac{1}{n}
\min\left\{\ln \frac{1}{P_{1}}, \ln\frac{1}{P_{20}}\right\} -
\frac{2}{n} \geq \\
\geq \min\left\{\gamma E_{\rm low}(R/\gamma,p,2),\gamma
E_{\rm low}(R/\gamma,p) + (1-\gamma)E_{2}(p)\right\} + o(1)\,,
\end{gathered}
$$
where $E_{2}(p)$ is defined in (\ref{defE2}). Therefore
\begin{equation}\label{nonoise1}
\begin{gathered}
F_{1}(R,p) \geq \max_{0 \leq \gamma \leq 1} \min\left\{\gamma
E_{\rm low}(R/\gamma,p,2),\gamma E_{\rm low}(R/\gamma,p) +
(1-\gamma)E_{2}(p)\right\},
\end{gathered}
\end{equation}
where $E_{\rm low}(R,p,2)$ and $E_{\rm low}(R,p)$ are defined in
(\ref{Elow}) (see also \S\,6).

Note that the function $\gamma E_{\rm low}(R/\gamma,p,2)$ from the
right-hand side of (\ref{nonoise1}) monotonically increases in
$\gamma$. On the contrary, the function
$S(\gamma,R,p)= \gamma E_{\rm low}(R/\gamma,p) + (1-\gamma)E_{2}(p)$
monotonically decreases in $\gamma$. Indeed, denoting
$r = R/\gamma$ and omitting $p$, we have
$S'_{\gamma}(\gamma,R) = E_{\rm low}(r) -rE'_{\rm low}(r) - E_{2}$
and $S''_{\gamma r}(\gamma,R) = -rE''_{\rm low}(r) < 0$. Therefore
maximum over $R,\gamma$ of the value $S'_{\gamma}(\gamma,R)$ is
attained when $r \to 0$. Since $rE'_{\rm low}(r) \to 0,\,r \to 0$,
then we get $\max\limits_{R,\gamma} S'_{\gamma}(\gamma,R) =
E_{\rm low}(0) -E_{2} < 0$.

We consider only the case $R < R_{\rm crit}(p)$, i.e. when
$E_{\rm low}(R,p,2) > E_{\rm low}(R,p)$. For such $R$ the best is to
set $\gamma = \gamma_{0}$ such that $P_{1} = P_{20}$, i.e.
\begin{equation}\label{Pe3}
\begin{gathered}
\gamma_{0} E_{\rm low}(R/\gamma_{0},p,2) = \gamma_{0}
E_{\rm low}(R/\gamma_{0},p) + (1-\gamma_{0})E_{2}(p)\,.
\end{gathered}
\end{equation}
Both sides of (\ref{Pe3}) are continuous functions in $\gamma_{0}$.
The left-hand side of (\ref{Pe3}) monotonically increases in
$\gamma_{0}$, and the right-hand one monotonically decreases in
$\gamma_{0}$. With $\gamma_{0} = 1$ the left-hand side is greater
than its right-hand side, which equals $E_{\rm low}(R,p)$. On the
contrary, for $\gamma_{0} = R/R_{\rm crit}$ the right-hand side is
greater than the left-hand side. Then there exists the unique
$\gamma_{0} \in (R/R_{\rm crit},1)$ satisfying (\ref{Pe3}).
Therefore we get
\begin{equation}\label{Prop1a}
F_{1}(R,p) \geq \gamma_{0} E_{\rm low}(R/\gamma_{0},p) +
(1-\gamma_{0})E_{2}(p)> E_{\rm low}(R,p).
\end{equation}

We show that, in fact, $F_{1}(R,p)$ satisfies the stronger
inequality (\ref{Theor2}), although we know exactly only part of the
function $E(R,p), 0 < R < R_{\rm crit}(p)$ (see (\ref{main2})). If
we connect the points $E(0,p)$ and $E(R_{\rm crit},p)$ by the piece
of the straight line, then due to the ``straight-line bound''
\cite{SGB1}, for $0 \leq R \leq R_{\rm crit}$ the function $E(R,p)$
does not exceed that straight line. Therefore, if
$0 < R < R_{\rm crit}(p)$ and $0 < p < 1/2$ then the inequality
holds
$$
E(R,p) < E(0,p) - \frac{[E(0,p) - E(R_{\rm crit}(p),p)]R}
{R_{\rm crit}(p)}\,.
$$
Now, to establish the formula (\ref{Theor2}), it is sufficient to
check that for such $p,R$ the following strict inequality is valid
\begin{equation}\label{Prop1b}
\gamma_{0}E_{\rm ex}(R/\gamma_{0},p,2) > E(0,p) - \frac{[E(0,p) -
E(R_{\rm crit}(p),p)]R}{R_{\rm crit}(p)}\,.
\end{equation}
For that purpose it is convenient to introduce the parameter
$u = R/\gamma_{0},\,u \in (0,R_{\rm crit})$. Then we get the
parametric representation for $\gamma_{0} = \gamma_{0}(u,p)$ and
$R = R(u,p)$:
$$
\begin{gathered}
\gamma_{0} = \frac{E_{2}(p)}{E_{2}(p) + E_{\rm ex}(u,p,2) -
E_{\rm ex}(u,p)}\,, \qquad R = u\gamma_{0}.
\end{gathered}
$$
Then combining analytical and numerical methods, it is not difficult
to check validity of the inequality (\ref{Prop1b}).
It concludes proof of the Theorem 2.
$\qquad \qquad \blacktriangle$.

In Fig. 3 the plots of the functions $F_{1}(R,p)$ and
$E_{\rm ex}(R,p)$ for $p = 0.01$ ($R_{\rm crit} \approx 0.387$)
are shown.

To compare  the functions $F_{1}(R,p)$ and $E(R,p)$ consider

{E x a m p l e \,1}.  Let $p=(1-\varepsilon)/2,\,\varepsilon \to 0$.
Then
$$
C(p) = \frac{\epsilon^{2}}{2} + O(\varepsilon^{4}), \quad
R_{\rm crit}(p) = \frac{C(p)[1+o(1)]}{4}, \quad
R_{\rm min 2}(p) \leq R_{\rm min}(p) = O(C^{2}).
$$
Therefore when $p \to 1/2$ the expurgation bound, essentially, is
not applicable and we get the known results \cite{G2}
$$
E(R,p)[1+o(1)] = \left\{\begin{array}{cc}
  C/2 - R, & 0 \leq R \leq C/4, \\
  (\sqrt{C} - \sqrt{R})^{2}, & C/4 \leq R \leq C,
\end{array}
\right.
$$
and
$$
\begin{gathered}
E(R,p,2)[1+o(1)] \geq  E_{r}(R,p,2) =
\left\{\begin{array}{cc}
  2C/3 - 2R, & 0 \leq R \leq C/9, \\
  (\sqrt{C} - \sqrt{R})^{2}, & C/9 \leq R \leq C.
\end{array}
\right.
\end{gathered}
$$
From those formulas and (\ref{deft0}) we also have
\begin{equation}\label{to1}
\begin{gathered}
4\varepsilon t_{0}(R,p)[1+o(1)] = \left\{\begin{array}{cc}
  C-6R, & 0 \leq R \leq C/9, \\
  3(\sqrt{C} - 2\sqrt{R})^{2}, & C/9 \leq R \leq C/4, \\
  0, & C/4 \leq R \leq C.
\end{array}
\right.
\end{gathered}
\end{equation}

Consider the equation (\ref{Pe3}). For
$R < R_{\rm crit}(p) = C(p)[1+o(1)]/4$, there are possible two
cases: $R/\gamma_{0} \leq C/9$ and $C/9 \leq R/\gamma_{0} < C/4$.

1) Let $R/\gamma_{0} \leq C/9$. Then from (\ref{Pe3}) we get
$$
\gamma_{0} = \frac{6(R+C)}{7C}\,, \qquad
F_{1}(R,p)= \frac{4C-10R}{7}\,, \qquad R \leq \frac{2C}{19}\,,
$$
and
$$
\frac{F_{1}(R,p)}{E(R,p)} = \frac{8}{7} -\frac{4R}{7(C-2R)}\,,
\qquad R \leq \frac{2C}{19}\,.
$$
The ratio $F_{1}(R,p)/E(R,p)$ monotonically decreases from $8/7$
(for $R = 0$) down to $16/15$ (for $R = 2C/19$).

2) Let $C/9 \leq R/\gamma_{0} < C/4$. Then we get
$$
\sqrt{\gamma_{0}} = \frac{2\sqrt{R}+\sqrt{6C - 8R}}{3\sqrt{C}}\,,
\qquad \frac{2C}{19} \leq R < \frac{C}{4}\,,
$$
and
$$
\begin{gathered}
F_{1}(R,p) = \frac{1}{9}\left[6C - 7R - 2\sqrt{2R(3C - 4R)}\right].
\end{gathered}
$$
The ratio $F_{1}(R,p)/E(R,p)$ monotonically decreases from
$16/15$ (for $R = 2C/19$) down to $1$ (for $R = C/4$).

It is natural to expect that similar results will also hold in the
case of the noisy feedback channel ${\rm BSC}(p_{1})$, if $p_{1}$ is
sufficiently small.


\begin{center}
{\bf \S\,4. Channel with noisy feedback. Proof of Theorem 1}
\end{center}

In the noisy feedback case we will still use the transmission method
with one switching moment. But if we try to use exactly the same
method as in the noiseless feedback case, we will face with the
following problem. After phase I, the transmitter should find the
two most probable (for the receiver) codewords
$\mbox{\boldmath $x$}^{1},\mbox{\boldmath $x$}^{2}$. But with
relatively high probability, the second and the third ranked
codewords $\mbox{\boldmath $x$}^{2}$ and $\mbox{\boldmath $x$}^{3}$
will be approximately equiprobable, and therefore it will be
difficult to the transmitter to rank them correctly (due to noise in
the feedback). Fortunately, in that case (with high probability) the
most probable codeword $\mbox{\boldmath $x$}^{1}$ will be much more
probable than $\mbox{\boldmath $x$}^{2}$, and then (again with high
probability) $\mbox{\boldmath $x$}^{1}$ is the true codeword. We
use this observation as follows: if posterior probabilities of the
second $\mbox{\boldmath $x$}^{2}$ and the third
$\mbox{\boldmath $x$}^{3}$ ranked codewords are not very different,
the receiver makes a decision immediately after phase I (in favor of
the most probable codeword $\mbox{\boldmath $x$}^{1}$), and it
ignores all next signals from the transmitter on phase II.

As a result, we use the following transmission and decoding method.

\underline{Transmission}. We set a number $0 < \gamma < 1$.
On phase I, of
length $m = \gamma n$, we use a ``good'' code (it is explained
below). Let $\mbox{\boldmath $x$}_{\rm true}$ be the transmitted
codeword of length $m$, $\mbox{\boldmath $y$}$ be the received (by
the receiver) block, and $\mbox{\boldmath $x$}'$ be the received
(by the transmitter) block. The transmitter selects one more
codeword
$\mbox{\boldmath $x$}_{i} \neq \mbox{\boldmath $x$}_{\rm true}$,
closest to $\mbox{\boldmath $x$}'$. For example, the codeword
$\mbox{\boldmath $x$}_{1} \neq \mbox{\boldmath $x$}_{\rm true}$ is
chosen, if $d(\mbox{\boldmath $x$}_{1},\mbox{\boldmath $x$}') =
\min\limits_{\mbox{\boldmath $x$}_{i} \neq
\mbox{\boldmath $x$}_{\rm true}}
d(\mbox{\boldmath $x$}_{i},\mbox{\boldmath $x$}')$. As a result, the
transmitter builds a list of two messages: the true one
$\theta_{\rm true}$ and another message
$\theta_{i} \neq \theta_{\rm true}$, which looks most probable
among remaining ones.

A ``good'' code in use of length $m$ should have the following
properties: \\
1) Its decoding error probability $P_{e}$ satisfies the inequality
$P_{e} \leq e^{-E_{\rm low}(R,p)m}$; \\
2) Its list size $L = 2$ decoding error probability $P_{e}(2)$
satisfies similar inequality
$P_{e}(2) \leq e^{-E_{\rm low}(R,p,2)m}$; \\
3) The relations (\ref{Pe2}) and (\ref{P1}) hold for it.

Existence of such code is shown in \S\,6, slightly modifying
standard Gallager's arguments for expurgation bound \cite{G2, G1}.

On phase II (i.e. on $(\gamma n, n]$) the transmitter uses the two
opposite codewords of length $n-m = (1-\gamma)n$ (for example,
consisting of all zeros and all ones), in order to help the receiver
to decide between the true message $\theta_{\rm true}$ and another
most probable message $\theta_{i} \neq \theta_{\rm true}$.

This transmission method is a slight modification of the method
used in \cite{BY1}. It gives the same decoding error probability
exponent, but it is simpler for analysis.
If the true message $\theta_{\rm true}$ is not among the two most
probable messages for the receiver, then there will always be the
decoding error. A slight modification of the transmission method
from \cite{BY1} used here helps in the case when the true message
$\theta_{\rm true}$ is among the two most probable messages for the
receiver, but it is not such one for the transmitter.

\underline{Decoding}. We set a number $t > 0$. Arrange the Hamming
distances $\{d(\mbox{\boldmath $x$}_{i},\mbox{\boldmath $y$}),\,
i=1,\ldots,M\}$ after phase I in the increasing order, denoting
$$
d^{(1)} = \min_{i} d(\mbox{\boldmath $x$}_{i},\mbox{\boldmath $y$})
\leq  d^{(2)} \leq \ldots \leq d^{(M)} =
\max_{i} d(\mbox{\boldmath $x$}_{i},\mbox{\boldmath $y$}),
$$
(in case of tie we use any order). Let also
$\mbox{\boldmath $x$}^{1},\ldots,\mbox{\boldmath $x$}^{M}$ be the
corresponding ranking of codewords after phase I, i.e
$\mbox{\boldmath $x$}^{1}$ is the closest to $\mbox{\boldmath $y$}$
codeword, etc. Two cases are possible.

C a s e \,1. If $d^{(3)} \leq d^{(2)} + t\gamma n$, then the
receiver makes the decoding immediately after phase I (in favor
of the closest to $\mbox{\boldmath $y$}$ codeword
$\mbox{\boldmath $x$}^{1}$). Although the transmitter will still
send some signals on phase II, the receiver has already made its
decision.

C a s e \,2. If $d^{(3)} > d^{(2)} + t\gamma n$, then after
phase I the receiver selects two most probable messages
$\theta_{i},\theta_{j}$, and after transmission on phase II
(i.e. after moment $n$) makes a decision between those two remaining
messages $\theta_{i},\theta_{j}$ in favor of more probable of them
(based on all received on $[0,n]$ signals).

In the case 2 the transmitter and the receiver will perform in
coordination, if the lists of two messages build by each of them
coincide. Remind that the receiver's list always contains the true
message. Of course, those lists may be different (and then there
will be the decoding error), but probability of such event should
be sufficiently small (which will be secured below).

{\it Remarks} 5. a) In the case of noiseless feedback (i.e. when
$p_{1} = 0$) the strategy described reduces to the strategy from
\S\,3 if we set $t = 0$. \\
b) The strategy described can be improved by introducing an
additional parameter $\tau \geq 0$, such that if
$d^{(2)} \geq d^{(1)} + \tau \gamma n$
then the receiver also makes the decoding immediately after phase I
(in favor of the closest to $\mbox{\boldmath $y$}$ codeword
$\mbox{\boldmath $x$}^{1}$). But introduction of such parameter
leads to too bulky formulas.

To evaluate the decoding error probability $P_{\rm e}$, denote by
$P_{1}$ and $P_{2}$ the decoding error probabilities in the case 1
(i.e. after the moment $\gamma n$), and in the case 2 (i.e. after
the moment $n$), respectively. Then for $P_{\rm e}$ we have
\begin{equation}\label{PetotaMl2}
P_{\rm e} \leq P_{1} + P_{2}\,.
\end{equation}

We evaluate the probabilities $P_{1}, P_{2}$ in the right-hand side
of (\ref{PetotaMl2}). Denoting
$d_{i} = d(\mbox{\boldmath $x$}_{i},\mbox{\boldmath $y$}),\\
i=1,\ldots,M$, for $P_{1}$ we have
\begin{equation}\label{PeII}
\begin{gathered}
P_{1} \leq M^{-1}\sum_{k=1}^{M}{\bf P}(d_{k} \neq d^{(1)};
d_{k} \geq d^{(3)} - t\gamma n|\mbox{\boldmath $x$}_{k}).
\end{gathered}
\end{equation}

We show that there exists a code such that for $P_{1}$ we have
($n \to \infty$)
\begin{equation}\label{P1}
\begin{gathered}
\frac{1}{n}\ln \frac{1}{P_{1}} \geq \gamma E_{\rm low}(R/\gamma,p,2)
-\frac{t \gamma}{3}\ln\frac{q}{p} + o(1)\,.
\end{gathered}
\end{equation}
Indeed, using the inequality $\left(\sum a_{i}\right)^{1/\rho} \leq
\sum a_{i}^{1/\rho},\,\rho \geq 1$, we have
$$
\begin{gathered}
{\bf P}^{1/\rho}\left(d_{k} \neq d^{(1)};d_{k} \geq
d^{(3)} - t\gamma n|\mbox{\boldmath $x$}_{k}\right) \leq \\
\leq 2^{1/\rho}{\bf P}^{1/\rho}\left(d_{k} = d^{(2)} \geq
d^{(3)} - t\gamma n|\mbox{\boldmath $x$}_{k}\right) +
2^{1/\rho}{\bf P}^{1/\rho}\left(d_{k} \geq d^{(3)}|
\mbox{\boldmath $x$}_{k}\right) \leq \\
\leq 2^{1+1/\rho}\left(\frac{q}{p}\right)^{t \gamma n/(3\rho)}
\sum_{m_{1},m_{2}}
\left[\sum_{\mbox{\small\boldmath $y$}}
\left[P\left(\mbox{\boldmath $y$}|\mbox{\boldmath $x$}_{k}\right)
P\left(\mbox{\boldmath $y$}|\mbox{\boldmath $x$}_{m_{1}}\right)
P\left(\mbox{\boldmath $y$}|\mbox{\boldmath $x$}_{m_{2}}
\right)\right]^{1/3}\right]^{1/\rho},
\end{gathered}
$$
and then
$$
\begin{gathered}
\left[E{\bf P}^{1/\rho}\left(d_{k} \neq d^{(1)};d_{k} \geq
d^{(3)} - t\gamma n|\mbox{\boldmath $x$}_{k}\right)\right]^{\rho/n}
\leq 2^{(1+\rho)/n}\left(\frac{q}{p}\right)^{t\gamma/3}
e^{-\gamma E_{\rm ex}(R/\gamma,p,2)}.
\end{gathered}
$$
A similar inequality holds with $E_{\rm r}(R/\gamma,p,2)$ instead of
$E_{\rm ex}(R/\gamma,p,2)$. Therefore using the definition of
$E_{\rm low}(R/\gamma,p,2)$ (see (\ref{Elow})), we get the formula
(\ref{P1}).

For the value $P_{2}$ we have
\begin{equation}\label{Pe20}
\begin{gathered}
P_{2} \leq P_{20} + P_{2n}\,,
\end{gathered}
\end{equation}
where $P_{20}$ is the decoding error probability in the case 2
for the channel with noiseless feedback, and $P_{2n}$ is the
probability that the most probable codeword (excluding the true codeword
$\mbox{\boldmath $x$}_{\rm true}$) for the receiver is not such one
for the transmitter (moreover, the true codeword is among two most
probable codewords for the receiver).

For the value $P_{20}$ the formula (\ref{Pe2}) remains valid.

It remains us to evaluate $P_{2n}$. For that purpose consider the
ensemble of codes ${\cal C}$ in which each codeword is selected
independently with the probability $2^{-m}$ among all possible
binary vectors of length $m$. We are interested in the value
${\bf E}_{\cal C}P_{2n}^{1/\rho}({\cal C}),\,\rho \geq 1$, where
expectation is taken over randomly chosen codes ${\cal C}$. Clearly,
$$
{\bf P}\left(\mbox{\boldmath $y$}|\mbox{\boldmath $x$}_{\rm true}
\right) = q^{m}\binom{m}{d}\left(\frac{p}{q}\right)^{d}, \qquad
d = d\left(\mbox{\boldmath $x$}_{\rm true},\mbox{\boldmath $y$}
\right).
$$
For given blocks $\mbox{\boldmath $x$}_{\rm true}$ and
$\mbox{\boldmath $y$}$ all $(M-1)$ remaining codewords are
independently and equiprobably distributed among all $2^{m}$ binary
vectors of length $m$. The vector $\mbox{\boldmath $y$}$ is
transmited over the feedback channel ${\rm BSC}(p_{1})$ and the
transmitter receives the vector $\mbox{\boldmath $x$}'$.

Without loss of generality we assume that
$\mbox{\boldmath $x$}_{\rm true} = \mbox{\boldmath $x$}_{M}$.
For the received block $\mbox{\boldmath $y$}$
we arrange all remaining codewords
$\mbox{\boldmath $x$}_{1},\ldots,\mbox{\boldmath $x$}_{M-1}$ as
$\mbox{\boldmath $x$}^{1},\ldots,\mbox{\boldmath $x$}^{M-1}$, in
increasing by their distance
$d\left(\mbox{\boldmath $x$}^{i},\mbox{\boldmath $y$}\right)$ order,
i.e. $d\left(\mbox{\boldmath $x$}^{1},\mbox{\boldmath $y$}\right)$
is the minimal distance, etc. In the case 2 it is necessary to have
$d(\mbox{\boldmath $x$}^{i},\mbox{\boldmath $y$}) -
d(\mbox{\boldmath $x$}^{1},\mbox{\boldmath $y$}) \geq tm,\,
i=2,\ldots,M-1$ (otherwise, the case 1 occurs). Moreover, we may
assume that the distance
$d(\mbox{\boldmath $x$}^{1},\mbox{\boldmath $y$})$
satisfies the condition ($m \to \infty$)
\begin{equation}\label{condP2n1}
\begin{gathered}
d(\mbox{\boldmath $x$}^{1},\mbox{\boldmath $y$})/m \leq
\delta_{GV}(R/\gamma) - t + o(1), \qquad R > 0,
\end{gathered}
\end{equation}
which is equivalent to the inequality
$$
h\left\{d(\mbox{\boldmath $x$}^{1},\mbox{\boldmath $y$})/m +t
\right\} \leq \ln 2 - R/\gamma \,, \qquad
d(\mbox{\boldmath $x$}^{1},\mbox{\boldmath $y$})/m+t < 1/2\,.
$$

Indeed, blocks $\mbox{\boldmath $y$},\mbox{\boldmath $x$}_{1},
\ldots, \mbox{\boldmath $x$}_{M-1}$ are distributed independently
and equiprobably among all $2^{m}$ binary vectors of length $m$.
For $u \geq 0$ introduce the random event
$$
\begin{gathered}
{\cal A}(u) = \{d(\mbox{\boldmath $x$}^{1},\mbox{\boldmath $y$}) >
(u-t)m; \; d(\mbox{\boldmath $x$}^{2},\mbox{\boldmath $y$}) -
d(\mbox{\boldmath $x$}^{1},\mbox{\boldmath $y$}) \geq tm \}.
\end{gathered}
$$
Then
$$
\begin{gathered}
{\bf P}\{{\cal A}(u)\} \leq
(M-1){\bf P}\left\{d(\mbox{\boldmath $x$}_{1},\mbox{\boldmath $y$})
> (u-t)m \right\} \prod_{i=2}^{M-1}
{\bf P}\left\{d(\mbox{\boldmath $x$}_{i},\mbox{\boldmath $y$}) >
um \right\} = \\
= (M-1){\bf P}\{w(\mbox{\boldmath $x$}_{1}) > (u-t)m\}
{\bf P}^{M-2}\{w(\mbox{\boldmath $x$}_{2}) > um\} \leq \\
\leq (M-1) \left[1 - {\bf P}\left\{w(\mbox{\boldmath $x$}_{2}) \leq
um \right\}\right]^{M-2} \leq (M-1)
\exp\left\{-(M-2)P\left\{w(\mbox{\boldmath $x$}) \leq um\right\}
\right\},
\end{gathered}
$$
where the inequality $(1-a)^{b} \leq e^{-ab},\,b \geq 0$ was used.
Note that
$$
\begin{gathered}
P\left\{w(\mbox{\boldmath $x$}) \leq um\right\} \geq 2^{-m}
\binom{m}{um} \geq \frac{1}{(m+1)}2^{-m}e^{mh(u)},
\end{gathered}
$$
since \cite[формула (12.40)]{CT} for any $0 \leq k \leq n$ the
inequalities hold
$$
\frac{1}{n+1}2^{nh(k/n)} \leq \binom{n}{k} \leq 2^{nh(k/n)}.
$$
Therefore
$$
\begin{gathered}
{\bf P}\{{\cal A}(u)\} \leq \exp\left\{\frac{Rm}{\gamma} -
\frac{(M-2)}{M(m+1)}\,e^{[R/\gamma + h(u)- \ln 2]m}\right\}.
\end{gathered}
$$
We set $u$ such that $[R/\gamma + h(u)- \ln 2]m \geq 4\ln m$. Then
for sufficiently large $m$ we have
${\bf P}\{{\cal A}(u)\} \leq e^{-m^{2}}$, and we may neglect the
event of such small probability. Therefore the inequality
(\ref{condP2n1}) holds.

Assuming that
$\mbox{\boldmath $x$}_{\rm true} = \mbox{\boldmath $x$}_{M}$,
For given $\mbox{\boldmath $y$},\mbox{\boldmath $x$}',
\mbox{\boldmath $x$}_{M}$ and randomly (equiprobably) chosen
$\mbox{\boldmath $x$}_{1},\mbox{\boldmath $x$}_{2}$ introduce
the set
$$
\begin{gathered}
{\cal F}(\mbox{\boldmath $y$},\mbox{\boldmath $x$}',
\mbox{\boldmath $x$}_{M}) = \left\{\mbox{\boldmath $x$}_{1},
\mbox{\boldmath $x$}_{2}: \begin{array}{c}
d(\mbox{\boldmath $x$}_{1},\mbox{\boldmath $y$}) \leq
\delta_{GV}(R/\gamma)m - tm,  \;
d(\mbox{\boldmath $x$}_{1},\mbox{\boldmath $x$}') \geq
d(\mbox{\boldmath $x$}_{2},\mbox{\boldmath $x$}')
\end{array}\right\}.
\end{gathered}
$$
We are interested in the values
$P_{3} = {\bf P}\left\{{\cal F}(\mbox{\boldmath $y$},
\mbox{\boldmath $x$}',\mbox{\boldmath $x$}_{M})\big|
\mbox{\boldmath $y$},\mbox{\boldmath $x$}', \mbox{\boldmath $x$}_{M}
\right\}$ and ${\bf E}_{\mbox{\boldmath $y$},\mbox{\boldmath $x$}',
\mbox{\boldmath $x$}_{M}}P_{3}^{s},\,s \geq 0$.

{\it Remark} 6. In the definition of the set
${\cal F}(\mbox{\boldmath $y$},
\mbox{\boldmath $x$}',\mbox{\boldmath $x$}_{M})$ we might include
additional constraints: $d(\mbox{\boldmath $x$}_{2},
\mbox{\boldmath $y$}) \geq \delta_{GV}(R/\gamma)m;\,
d(\mbox{\boldmath $x$}_{2},\mbox{\boldmath $y$}) \geq
d(\mbox{\boldmath $x$}_{M},\mbox{\boldmath $y$})$. But
it seems that they do not improve the exponent of $P_{3}$.

Note that if $d(\mbox{\boldmath $y$},\mbox{\boldmath $y$}') \leq tm$
then $P_{2n} = P_{3} = 0$. Moreover, if $p_{1} < t$ then
\begin{equation}\label{upP2n}
P_{2n} \leq P\left\{d(\mbox{\boldmath $y$},\mbox{\boldmath $x$}')
\geq tm\right\} \leq e^{-mD(t\|p_{1})}.
\end{equation}
If $d(\mbox{\boldmath $y$},\mbox{\boldmath $x$}') > tm$, then for
any nonnegative $\alpha,\varphi$
$$
\begin{gathered}
P_{3} \leq {\bf E}_{\mbox{\boldmath $x$}_{1},
\mbox{\boldmath $x$}_{2}}\big\{e^{\alpha[(\delta - t)m -
d(\mbox{\boldmath $x$}_{1},\mbox{\boldmath $y$})] +
\varphi[d(\mbox{\boldmath $x$}_{1},\mbox{\boldmath $x$}')
- d(\mbox{\boldmath $x$}_{2},\mbox{\boldmath $x$}')]}\big|
\mbox{\boldmath $y$},\mbox{\boldmath $x$}',
\mbox{\boldmath $x$}_{M}\big\} = \\
= e^{\alpha(\delta - t)m}{\bf E}_{\mbox{\boldmath $x$}_{1},
\mbox{\boldmath $x$}_{2}}\big\{
e^{-\alpha d(\mbox{\boldmath $x$}_{1},\mbox{\boldmath $y$}) +
\varphi[d(\mbox{\boldmath $x$}_{1},
\mbox{\boldmath $x$}') - d(\mbox{\boldmath $x$}_{2},
\mbox{\boldmath $x$}')]}\big|\mbox{\boldmath $y$},
\mbox{\boldmath $x$}',\mbox{\boldmath $x$}_{M}\big\}.
\end{gathered}
$$
For any $a,b$ and equiprobable $\mbox{\boldmath $x$}$
$$
\begin{gathered}
{\bf E}_{\mbox{\boldmath $x$}}\left[
e^{ad(\mbox{\boldmath $x$},\mbox{\boldmath $y$}) +
bd(\mbox{\boldmath $x$},\mbox{\boldmath $x$}')}\big|
\mbox{\boldmath $y$},\mbox{\boldmath $x$}'\right] =
2^{-m}\left(1+e^{a+b}\right)^{m}\left(\frac{e^{a} + e^{b}}
{1+e^{a+b}}\right)^{d(\mbox{\boldmath $y$},\mbox{\boldmath $x$}')}.
\end{gathered}
$$
Then when
$d(\mbox{\boldmath $y$},\mbox{\boldmath $x$}') > tm$, we have
$$
\begin{gathered}
P_{3} \leq 2^{-2m}e^{\alpha(\delta - t)m}
\left(1+e^{\varphi - \alpha}\right)^{m}
\left(1+e^{-\varphi}\right)^{m}
\left[\frac{e^{-\alpha} + e^{\varphi}}{1+e^{\varphi - \alpha}}
\right]^{d(\mbox{\boldmath $y$},\mbox{\boldmath $x$}')}.
\end{gathered}
$$
Since ${\bf E}b^{d(\mbox{\boldmath $y$},\mbox{\boldmath $x$}')} =
(q_{1} + p_{1}b)^{m}$, then
$$
\begin{gathered}
\left\{{\bf E}\left[b^{d(\mbox{\boldmath $y$},
\mbox{\boldmath $x$}')}; d(\mbox{\boldmath $y$},
\mbox{\boldmath $x$}') > tm\right] \right\}^{1/m} \leq
\left\{\min_{\mu \geq 0}
{\bf E}b^{d(\mbox{\boldmath $y$},\mbox{\boldmath $x$}') + \mu
[d(\mbox{\boldmath $y$},\mbox{\boldmath $x$}') - tm]}
\right\}^{1/m} = \\
= \min_{\mu \geq 0}\left\{b^{-\mu t}(q_{1} + p_{1}b^{1+\mu})
\right\}.
\end{gathered}
$$
Note that ($b \geq 1$)
\begin{equation}\label{case1}
\begin{gathered}
\min_{\mu \geq 0}\left\{b^{-\mu t}\left(z_{1} + b^{1+\mu}\right)
\right\} = e^{f_{4}(b,t,p_{1})}, \\
f_{4}(b,t,p_{1}) = \left\{\begin{array}{cc}
h(t) + (1-t)\ln z_{1} + t\ln b, &
\ln(tz_{1}/(1-t)) \geq \ln b, \\
\ln\left(z_{1} + b\right), &
\ln(tz_{1}/(1-t)) \leq \ln b,
\end{array}
\right.
\end{gathered}
\end{equation}
where minimum is attained when
$$
\mu = \mu_{0}= \left[\frac{\ln(tz_{1}/(1-t))}{\ln b}-1\right]_{+}.
$$
Therefore for $b_{1} \geq 1$ we have
$$
\begin{gathered}
\left({\bf E}_{\mbox{\boldmath $y$},\mbox{\boldmath $x$}',
\mbox{\boldmath $x$}_{M}}P_{3}^{s}\right)^{1/m} \leq
2^{-2s}e^{\alpha(\delta - t)s}e^{-(\alpha + \varphi)s +
f_{4}(b_{1}^{s},t,p_{1})}\left(e^{\alpha}+e^{\varphi}\right)^{s}
\left(e^{\varphi}+ 1\right)^{s},
\end{gathered}
$$
where
$$
\begin{gathered}
b_{1} = \frac{1 +e^{\varphi + \alpha}}{e^{\alpha}+e^{\varphi}}.
\end{gathered}
$$
We should minimize that expression over nonnegative
$\alpha,\varphi$. We have
$$
\begin{gathered}
{\bf E}e^{ad(\mbox{\boldmath $x$}_{M},\mbox{\boldmath $y$})} =
\left(q + pe^{a}\right)^{m}, \qquad
{\bf E}b^{d(\mbox{\boldmath $y$},\mbox{\boldmath $x$}')} =
(q_{1} + p_{1}b)^{m}.
\end{gathered}
$$
Denote
\begin{equation}\label{zz1b1}
\begin{gathered}
z = \frac{q}{p}, \qquad z_{1} = \frac{q_{1}}{p_{1}},
\end{gathered}
\end{equation}
and note that
$$
\begin{gathered}
b_{1} -1 \sim \left(e^{\varphi} - e^{-\varphi}\right)
\left(1 - e^{-\alpha}\right) \geq 0\,.
\end{gathered}
$$
Then
\begin{equation}\label{EF11}
\begin{gathered}
\left({\bf E}_{\mbox{\boldmath $y$},\mbox{\boldmath $x$}',
\mbox{\boldmath $x$}_{M}}P_{3}^{s}\right)^{1/m} \leq 2^{-2s}p_{1}
e^{-[\alpha(1-\delta + t) +\varphi]s +
f_{4}(b_{1}^{s},t,p_{1})}\left(e^{\alpha}+e^{\varphi}\right)^{s}
\left(e^{\varphi}+ 1\right)^{s}.
\end{gathered}
\end{equation}

We apply the random coding with expurgation method,
using the inequality \\ $\left(\sum a_{i}\right)^{1/\rho} \leq
\sum a_{i}^{1/\rho},\,\rho \geq 1$. We have
$$
\begin{gathered}
{\bf E}_{\cal C}P_{2n}^{1/\rho}({\cal C}) \leq
M^{2}{\bf E}_{\mbox{\boldmath $y$},\mbox{\boldmath $x$}',
\mbox{\boldmath $x$}_{M}}{\bf P}^{1/\rho}\left\{{\cal F}
(\mbox{\boldmath $y$},\mbox{\boldmath $x$}',
\mbox{\boldmath $x$}_{M})\big|\mbox{\boldmath $y$},
\mbox{\boldmath $x$}', \mbox{\boldmath $x$}_{M}\right\} =
M^{2}E_{\mbox{\boldmath $y$},\mbox{\boldmath $x$}',
\mbox{\boldmath $x$}_{M}}P_{3}^{1/\rho}
\end{gathered}
$$
and then from (\ref{EF11}) we get ($\rho = 1/s \geq 1$)
$$
\begin{gathered}
\left[{\bf E}_{\cal C}P_{2n}^{1/\rho}({\cal C})\right]^{\rho/m} \leq
e^{2R\rho/\gamma}2^{-2}p_{1}^{\rho}
e^{\rho f_{4}(b_{1}^{s},t,p_{1}) -\alpha(1-\delta + t) -
\varphi}\left(e^{\alpha}+e^{\varphi}\right)
\left(e^{\varphi}+ 1\right).
\end{gathered}
$$
To avoid bulky formulas, we choose the parameters such that the
inequality holds (see (\ref{case1}))
\begin{equation}\label{cond31}
\rho \ln(tz_{1}/(1-t)) \geq \ln b_{1}.
\end{equation}
Then
$$
\begin{gathered}
\left[{\bf E}_{\cal C}P_{2n}^{1/\rho}({\cal C})\right]^{\rho/m} \leq
2^{-2}e^{G\rho + F_{2}}, \qquad b_{1} = \frac{1+cd}{c+d}, \\
G = 2R/\gamma + h(t) +
\ln\left[p_{1}^{t}q_{1}^{1-t}\right] =
2R/\gamma - D\left(t\|p_{1}\right), \\
F_{2} = -(1-\delta+t)\ln d - \ln c
+ t\ln(1+dc) + \ln(1+c) + (1-t)\ln(d+c),
\end{gathered}
$$
and we should minimize $F_{2}$ over $c,d \geq 1$.

Note that $F_{2}$ does not depend on $\rho$. If $G < 0$ then the
best is $\rho \to \infty$. Since \\
$\left[{\bf E}_{\cal C}P_{2n}^{1/\rho}({\cal C})\right]^{\rho/m}
\to 0,\,\rho \to \infty$, we may assume that $P_{2n} = 0$.
If $G \geq 0$ then the best is $\rho = 1$ (and then it is better to
use simply the random coding method). In both cases we need the
condition (\ref{cond31}) be satisfied.

If $\rho \to \infty$ then the inequality (\ref{cond31}) is
equivalent to the condition $tz_{1}/(1-t) > 1$, i.e. $p_{1} < t$.
We set $t > p_{1}$ such that
$2R/\gamma - D\left(t\,\|\,p_{1}\right) < 0$. Then
$G < 0,\,P_{2n} = 0$, and from (\ref{P1}), (\ref{Pe2}) we get
\begin{equation}\label{noise1}
\begin{gathered}
F_{1}(R,p,p_{1}) \geq \max_{\gamma,t > p_{1}}\min\bigg\{\gamma
E_{\rm low}(R/\gamma,p,2) -\frac{t\gamma}{3}\ln\frac{q}{p},
\gamma E_{\rm low}(R/\gamma,p) + (1-\gamma)E_{2}(p)\bigg\}.
\end{gathered}
\end{equation}

Using $t = t_{1}(R,p_{1}) \geq p_{1}$ (see (\ref{deft_{1}}))
we get from (\ref{noise1})
\begin{equation}\label{noise2}
\begin{gathered}
F_{1}(R,p,p_{1}) \geq \max_{\gamma}\min\bigg\{\gamma
E_{\rm low}(R/\gamma,p,2) -\frac{\gamma t_{1}(R/\gamma,p_{1})}{3}
\ln\frac{q}{p}, \\
\gamma E_{\rm low}(R/\gamma,p) + (1-\gamma)E_{2}(p)\bigg\},
\end{gathered}
\end{equation}
from which the formulas (\ref{Theor1}), (\ref{Theor1a}) and the
Theorem 1 follow. $\qquad \blacktriangle$

{\it Remark} 7. Note that if $p_{1} \to 0$, then $t_{1} \to 0$
and the relation (\ref{noise2}) transfers to the similar relation
(\ref{nonoise1}) for the channel with noiseless feedback.

To find the function $p_{0}(R,p)$ of the critical noise level in the
feedback channel we set $\gamma \to 1$. Then $p_{0} = p_{0}(R,p)$
is defined by the system of equations
$$
\begin{gathered}
E_{\rm low}(R,p,2) -\frac{t}{3}\ln\frac{q}{p} =
E_{\rm low}(R,p), \\
D\left(t\,\|\,p_{0}\right) = 2R.
\end{gathered}
$$
In other words, $t_{0}(R,p)$ and $p_{0}(R,p) \leq t_{0}(R,p)$ are
defined by the formulas (\ref{deft0}) and (\ref{defp_{1}}),
respectively.

\begin{center}
{\bf \S\,5. When noisy feedback behaves like noiseless ?}
\end{center}

How small should be $p_{1}$ in order to have the error exponent
$F_{1}(R,p,p_{1})$ close to the similar exponent $F_{1}(R,p)$ for
noiseless feedback ? More exactly, when for a given
$\alpha \in (0,1)$ the inequality holds
$F_{1}(R,p,p_{1}) - E(R,p) \leq (1-\alpha)[F_{1}(R,p) - E(R,p)]$ ?

We give a simple estimate for such $p_{1}$, considering only the
case $R = 0$. For the optimal $\gamma = \gamma_{0}$ from
(\ref{Theor1}), (\ref{Theor1a}) we have ($E_{2}(p) = 2E(0,p)$)
$$
\begin{gathered}
\gamma_{0} = \frac{2E(0,p)}{E(0,p,2) + E(0,p)-p_{1}\ln(q/p)/3}
\end{gathered}
$$
and then
$$
\begin{gathered}
F_{1}(0,p,p_{1}) = \frac{2E(0,p)[E(0,p,2) - p_{1}\ln(q/p)/3]}
{E(0,p,2) + E(0,p)-p_{1}\ln(q/p)/3}, \\
F_{1}(0,p,p_{1}) - E(0,p) = \frac{E(0,p)[E(0,p,2) -E(0,p) -
p_{1}\ln(q/p)/3]}{E(0,p,2) + E(0,p)-p_{1}\ln(q/p)/3}.
\end{gathered}
$$
Now in order to have
$$
F_{1}(0,p,p_{1}) - E(0,p) \geq (1- \alpha)[F_{1}(0,p) - E(0,p)],
$$
it is sufficient to have
$$
p_{1} \leq \frac{3\alpha \left[E^{2}(0,p,2) -E^{2}(0,p)\right]}
{[\alpha E(0,p,2) + (2-\alpha)E(0,p)]\ln(q/p)}.
$$
Since $E(0,p,2) \geq E(0,p)$, without much loss, we may replace the
last inequality by a stronger one:
$$
p_{1} \leq \frac{3\alpha \left[E(0,p,2) - E(0,p)\right]}
{\ln(q/p)} = p_{11}(p,\alpha).
$$
On Fig. 4 the plot of the function $p_{11}(p,0.1)$ is given.

{E x a m p l e \,2}. Consider the case
$p =(1-\varepsilon)/2,\,\varepsilon \to 0$. Then
$C(p) \approx \epsilon^{2}/2$ and
$E(0,p,2) \approx 2C/3,\,E(0,p) \approx C/2$. As a result, we get
$$
p_{11}(p,\alpha) = \frac{\alpha (1-2p)[1+o(1)]}{8}, \qquad
p \to 1/2.
$$
In other words, if the forward ${\rm BSC}(p)$ is very bad, then in
order to improve its error exponent we need a very good feedback
channel ${\rm BSC}(p_{1})$.


\begin{center}
{\bf \S\,6. Auxiliary formulas and results}
\end{center}

{\bf Lower bounds for the decoding error exponents}.
All formulas below are derived following Gallager's technique
\cite{G2, G1}.

1) {\it Random coding} bounds:
\begin{equation}\label{lowranL}
E(R,p,L)\geq E_{\rm r}(R,p,L), \qquad R \geq 0.
\end{equation}
Moreover ($R_{{\rm crit}, L}(p)$ определено в (\ref{Rcrit})),
\begin{equation}\label{sphere1}
\begin{gathered}
E(R,p,L) = E_{\rm r}(R,p,L) = E_{\rm sp}(R,p), \qquad
R_{{\rm crit}, L}(p) \leq R \leq C(p),
\end{gathered}
\end{equation}
and for $R \leq R_{{\rm crit}, L}(p)$ we have
\begin{equation}\label{randomcodL}
E(R,p,L) \geq E_{\rm r}(R,p,L) = L(\ln 2 - R) -
(1+L)\ln\left[p^{1/(1+L)} +q^{1/(1+L)}\right].
\end{equation}
Since $R_{{\rm crit}, L}(p) \to 0,\,L \to \infty$, then
$E(R,p,L)\to E_{\rm sp}(R,p),\,L \to \infty$ for any $R \geq 0$.

2) {\it Random coding with expurgation bound}:
\begin{equation}\label{lowexpL}
E(R,p,L)\geq E_{\rm ex}(R,p,L) =
\max_{\rho \geq 1}\left\{- \rho LR -\rho \ln f(p,L,\rho)\right\},
\qquad R \geq 0,
\end{equation}
where
$$
\begin{gathered}
f(p,L,\rho) = 2^{-(L+1)}\left\{2 + \sum_{i=1}^{L}\binom{L+1}{i}
a_{i}^{1/\rho}\right\}, \\
a_{i} = p\left(\frac{q}{p}\right)^{i/(L+1)} +
q\left(\frac{p}{q}\right)^{i/(L+1)}.
\end{gathered}
$$
The bound (\ref{lowexpL}) improves the random coding bound
(\ref{randomcodL}) for $0 \leq R < R_{{\rm min}, L}(p)$ (see
(\ref{RminL}), but it does not give $E_{\rm sp}(R,p)$.
Note also that
\begin{equation}\label{derG1L}
\begin{gathered}
f(p,L,\rho) = {\bf E}\sum_{m,m_{1},\ldots,m_{L}}
\left[\sum_{\mbox{\small\boldmath $y$}}\left[
{\bf P}\left(\mbox{\boldmath $y$}|\mbox{\boldmath $x$}_{m}\right)
{\bf P}\left(\mbox{\boldmath $y$}|\mbox{\boldmath $x$}_{m_{1}}
\right)\ldots {\bf P}\left(\mbox{\boldmath $y$}|
\mbox{\boldmath $x$}_{m_{L}}\right)\right]^{1/(L+1)}
\right]^{1/\rho},
\end{gathered}
\end{equation}
where all components of each codeword $\mbox{\boldmath $x$}_{i}$
are chosen independently and equiprobably from $0$ and $1$.

In particular,
$$
\begin{gathered}
E_{\rm ex}(R,p) = E_{\rm ex}(R,p,1) = \max_{\rho \geq 1}
\left\{\rho\ln 2 - \rho R - \rho \ln\left[1+
\left(2\sqrt{pq}\right)^{1/\rho}\right]\right\}, \\
E_{\rm ex}(R,p,2) = \max_{\rho \geq 1}\left\{\rho\ln 4 - 2\rho R -
\rho \ln \left[1+3\left(p^{1/3}q^{2/3} +
p^{2/3}q^{1/3}\right)^{1/\rho}\right]\right\}.
\end{gathered}
$$

The functions $E(R,p,L),E_{\rm r}(R,p,L)$ and $E_{\rm ex}(R,p,L)$
does not decreases on $L$. In particular,
\begin{equation}\label{idea2}
E_{\rm ex}(R,p) < E_{\rm ex}(R,p,2), \qquad R < R_{\rm crit}(p).
\end{equation}

In order to get a more convenient representation for the functions
$E_{\rm ex}(R,p)$ and \\ $E_{\rm ex}(R,p,L)$, introduce rates
\begin{equation}\label{RminL}
\begin{gathered}
R_{{\rm min}, L}(p) =
\ln 2 - \frac{(L+1)}{L}\ln\left[p^{1/(L+1)} +q^{1/(L+1)}\right] -
\dfrac{\sum_{i=1}^{L}\binom{L+1}{i}a_{i}\ln a_{i}}
{2L\left[p^{1/(L+1)} + q^{1/(L+1)}\right]^{L+1}}.
\end{gathered}
\end{equation}
The function $R_{{\rm min}, L}(p)$ monotonically decreases on $L$
and $R_{{\rm min}, L}(p) < R_{{\rm crit}, L}(p)$, if
$L \geq 1$ and $0 < p < 1/2$. In particular,
\begin{equation}\label{Rmin12}
\begin{gathered}
R_{\rm min}(p) = R_{{\rm min}, 1}(p) = \ln 2 -
h\left(\frac{2\sqrt{pq}}{1+2\sqrt{pq}}\right), \\
R_{{\rm min}, 2}(p) = \ln 2 - \frac{1}{2}\left[\ln(1+3a_{1}) -
\frac{3a_{1}\ln a_{1}}{1+3a_{1}}\right], \qquad
a_{1} = p^{1/3}q^{2/3} + p^{2/3}q^{1/3}.
\end{gathered}
\end{equation}
We also have
$R_{{\rm min}, 2}(p) < R_{{\rm min}, 1}(p) < R_{\rm crit}(p),\,
0 < p < 1/2$.

Now
$$
\begin{gathered}
E_{\rm ex}(R,p,L) < E_{\rm r}(R,p,L) = E_{\rm sp}(R,p) \qquad
R > R_{{\rm crit}, L}(p), \\
E_{\rm ex}(R,p,L) = E_{\rm r}(R,p,L), \qquad
R_{{\rm min}, L}(p) \leq R \leq R_{{\rm crit}, L}(p), \\
E_{\rm ex}(R,p,L) > E_{\rm r}(R,p,L), \qquad
0 \leq R < R_{{\rm min}, L}(p).
\end{gathered}
$$
Moreover,
\begin{equation}\label{L1low}
\begin{gathered}
E_{\rm ex}(R,p) =  \frac{\delta_{GV}(R)}{2}\ln\dfrac{1}{4pq}\,,
\qquad 0 \leq R \leq R_{\rm min}(p).
\end{gathered}
\end{equation}
Note also that $0 \leq R \leq R_{\rm min}(p)$ corresponds to the
case $\delta_{GV}(R) \geq (2\sqrt{pq})/(1+2\sqrt{pq})$.

If $L = 2$, the
$$
E_{\rm ex}(R,p,2) = -v\ln a_{1}\,, \qquad
0 \leq R \leq R_{{\rm min}, 2}(p),
$$
where $a_{1}$ is defined in в (\ref{Rmin12}), and $v$ is the unique
root of the equation
$$
\ln 4 - h(v) - v\ln 3 = 2R\,, \qquad 0 \leq v < \frac{3}{4}\,.
$$
In particular,
\begin{equation}\label{Eex0}
\begin{gathered}
E_{\rm ex}(0,p) = E(0,p) =\frac{1}{2}\ln\frac{1}{2\sqrt{pq}}\,, \\
E_{\rm ex}(0,p,2) = E(0,p,2) = -\frac{3}{4}\ln\left(p^{1/3}q^{2/3} +
p^{2/3}q^{1/3}\right),
\end{gathered}
\end{equation}
(the second relation is established in \cite{Blin2}).

{\bf Existence of code with given properties}.
We are interested in a code ${\cal C}$ such that each
its codeword has certain properties
${\cal A}_{1},{\cal A}_{2},\ldots$. For that purpose we use the
following result which is a natural modification of the cute
Lemma 5.7 from \cite{G1}.

Assume that we choose randomly (in arbitrary way) a code ${\cal C}$
with $M'$ codewords $\mbox{\boldmath $x$}_{m}$, and for each
$\mbox{\boldmath $x$}_{m},\,m = 1,\ldots,M'$ we have
\begin{equation}\label{Lem1}
{\bf P}_{\mbox{\tiny over codes}}\left\{\mbox{\boldmath $x$}_{m}
\mbox{ does not have property }{\cal A}\right\} \leq 1/2\,.
\end{equation}

{L e m m a}. {\it If the condition} (\ref{Lem1}) {\it is satisfied
then there exists a code in the ensemble of codes with $M' = 2M-1$
codewords for which, at least, for $M$ its codewords the property
${\cal A}$ is fulfilled}.

P r o o f\, remains the same as in \cite[Lemma 5.7]{G1} (it is the
changing of the summation order in the corresponding double sum).
$\qquad \blacktriangle$

If there are, say, four properties
${\cal A}_{1},\ldots,{\cal A}_{4}$, then assume that for each
$\mbox{\boldmath $x$}_{m}$, $m = 1,\ldots,M'$, we have
\begin{equation}\label{Cor1}
\begin{gathered}
{\bf P}_{\mbox{\tiny over codes}}\left\{\mbox{\boldmath $x$}_{m}
\mbox{ does not have property }{\cal A}_{i}\right\} \leq 1/8\,,
\qquad i=1,\ldots,4\,.
\end{gathered}
\end{equation}

{C o r o l l a r y \,2}. {\it If the condition} (\ref{Cor1}) {\it is
satisfied, then there exists a code in the ensemble of codes with
$M' = 2M-1$ codewords for which, at least, for $M$ its codewords
all four properties ${\cal A}_{i},\,i=1,\ldots,4$ are fulfilled}.

In our case the property ${\cal A}_{1}$ means that the codeword
$\mbox{\boldmath $x$}_{m}$ has small decoding error probability;
${\cal A}_{2}$ means that $\mbox{\boldmath $x$}_{m}$ has small list
size $L = 2$ decoding error probability; ${\cal A}_{3},{\cal A}_{4}$
mean that for the codeword $\mbox{\boldmath $x$}_{m}$ the relations
(\ref{Pe2}) and (\ref{P1}), respectively, hold.

{\bf Proof of the formula (\ref{Pe2})}.
Consider a code ${\cal C}$ with $M$ codewords
$\mbox{\boldmath $x$}_{1},\ldots,\mbox{\boldmath $x$}_{M}$ of length
$n+k$. Each codeword $\mbox{\boldmath $x$}_{i}$ has the form
$\mbox{\boldmath $x$}_{i} = (\mbox{\boldmath $x$}_{i}',
\mbox{\boldmath $x$}_{i}'')$, where $\mbox{\boldmath $x$}_{i}'$ has
length $n$ and $\mbox{\boldmath $x$}_{i}''$ has length $k$. We
suppose that the parts $\{\mbox{\boldmath $x$}_{i}''\}$ are given,
while the parts $\{\mbox{\boldmath $x$}_{i}'\}$ are chosen randomly
(in some way). We also assume that
\begin{equation}\label{mind1}
\min_{i \neq j}d\left(\mbox{\boldmath $x$}_{i}'',
\mbox{\boldmath $x$}_{j}''\right) = \delta k \,.
\end{equation}
Using maximum likelihood decoding, denote by $P_{e,m}$ the
conditional decoding error \\ probability provided the codeword
$\mbox{\boldmath $x$}_{m}$ was transmitted. An output block
$\mbox{\boldmath $y$}$ has the form $\mbox{\boldmath $y$} =
(\mbox{\boldmath $y$}',\mbox{\boldmath $y$}'')$, where
$\mbox{\boldmath $y$}',\mbox{\boldmath $y$}''$ have length $n$ and
$k$, respectively. Then \\
${\bf P}\left(\mbox{\boldmath $y$}|\mbox{\boldmath $x$}_{m}\right) =
{\bf P}\left(\mbox{\boldmath $y$}'|\mbox{\boldmath $x$}_{m}'\right)
{\bf P}\left(\mbox{\boldmath $y$}''|
\mbox{\boldmath $x$}_{m}''\right)$.
Using the inequality
$\left(\sum a_{i}\right)^{s} \leq \sum a_{i}^{s},\,0 \leq s \leq 1$,
and the formula
$$
\sum_{\mbox{\small\boldmath $y$}'}
\sqrt{{\bf P}\left(\mbox{\boldmath $y$}'|
\mbox{\boldmath $x$}_{m}'\right)
{\bf P}\left(\mbox{\boldmath $y$}'|
\mbox{\boldmath $x$}_{m'}'\right)} =
\left(4pq\right)^{d\left(\mbox{\small\boldmath $x$}_{m}',
\mbox{\small\boldmath $x$}_{m'}'\right)/2},
$$
we have
\begin{equation}\label{derG1}
\begin{gathered}
P_{e,m}^{s} \leq \sum_{m' \neq m}
\left[\sum_{\mbox{\small\boldmath $y$}}
\sqrt{{\bf P}\left(\mbox{\boldmath $y$}|
\mbox{\boldmath $x$}_{m}\right)
{\bf P}\left(\mbox{\boldmath $y$}|
\mbox{\boldmath $x$}_{m'}\right)}\right]^{s} = \\
= \sum_{m' \neq m}\left[\sum_{\mbox{\small\boldmath $y$}'}
\sqrt{{\bf P}\left(\mbox{\boldmath $y$}'|
\mbox{\boldmath $x$}_{m}'\right)
{\bf P}\left(\mbox{\boldmath $y$}'|
\mbox{\boldmath $x$}_{m'}'\right)}\right]^{s}
\left[\sum_{\mbox{\small\boldmath $y$}''}\sqrt{
{\bf P}\left(\mbox{\boldmath $y$}''|
\mbox{\boldmath $x$}_{m}''\right)
{\bf P}\left(\mbox{\boldmath $y$}''\mbox{\boldmath $x$}_{m'}''
\right)}\right]^{s} \leq \\
\leq \sum_{m' \neq m}\left[\sum_{\mbox{\small\boldmath $y$}'}
\sqrt{{\bf P}\left(\mbox{\boldmath $y$}'|
\mbox{\boldmath $x$}_{m}'\right)
{\bf P}\left(\mbox{\boldmath $y$}'|
\mbox{\boldmath $x$}_{m'}'\right)}
\right]^{s} \left[\max_{m_{1} \neq m_{2}}
\left(2\sqrt{pq}\right)^{d\left(\mbox{\boldmath $x$}_{m_{1}}'',
\mbox{\boldmath $x$}_{m_{2}}''\right)}\right]^{s} =  \\
= \left(2\sqrt{pq}\right)^{\delta sk}
 \sum_{m' \neq m}\left[\sum_{\mbox{\small\boldmath $y$}'}
\sqrt{{\bf P}\left(\mbox{\boldmath $y$}'|
\mbox{\boldmath $x$}_{m}'\right)
{\bf P}\left(\mbox{\boldmath $y$}'|
\mbox{\boldmath $x$}_{m'}'\right)}\right]^{s} =
\left(2\sqrt{pq}\right)^{\delta sk}\sum_{m' \neq m}
\left(4pq\right)^{sd\left(\mbox{\small\boldmath $x$}_{m}',
\mbox{\small\boldmath $x$}_{m'}'\right)/2}.
\end{gathered}
\end{equation}
Consider an ensemble of codes in which each codeword
$\mbox{\boldmath $x$}_{m}'$ is selected independently with the
probability $2^{-n}$ among all possible binary vectors of length
$n$. Since
$$
\begin{gathered}
{\bf E}z^{d\left(\mbox{\small\boldmath $x$}_{m}',
\mbox{\small\boldmath $x$}_{m'}'\right)} =
{\bf E}z^{w\left(\mbox{\small\boldmath $x$}_{m}'\right)} =
\left(\frac{1+z}{2}\right)^{n}\,,
\end{gathered}
$$
we get
$$
\begin{gathered}
\left({\bf E} P_{e,m}^{s}\right)^{1/s} \leq
\left(2\sqrt{pq}\right)^{\delta k}\left\{e^{R}2^{-1}
\left[1+ \left(2\sqrt{pq}\right)^{s}\right]\right\}^{n/s}.
\end{gathered}
$$
Further derivation follows Theorem 5.7.1 from \cite{G1}. As a result,
defining $\rho = 1/s,\,\rho \geq 1$, we get that there exists a code
with $M$ codewords such that for any $m = 1,\ldots,M$ we have
$$
\begin{gathered}
\frac{1}{n}\ln\frac{1}{P_{e,m}} \geq \frac{\delta k}{n}
\ln\frac{1}{2\sqrt{pq}} + \max_{\rho \geq 1}\left\{\rho\ln 2 -
\rho R -
\rho \ln \left[1+ \left(2\sqrt{pq}\right)^{1/\rho}\right]\right\}.
\end{gathered}
$$
From that relation the formula (\ref{Pe2}) follows.
$\qquad \blacktriangle$

\medskip

The authors wish to thank the University of Tokyo for supporting
this joint research.


\vspace{5mm}

\begin{flushleft}
{\small {\it Burnashev Marat Valievich} \\
Institute for Information Transmission Problems RAS \\
 {\tt burn@iitp.ru}} \\
 {\small {\it Yamamoto Hirosuke} \\
The University of Tokyo, Japan \\
 {\tt hirosuke@ieee.org}}
\end{flushleft}%

\newpage

\includegraphics[width=0.9\hsize,height=0.8\hsize]{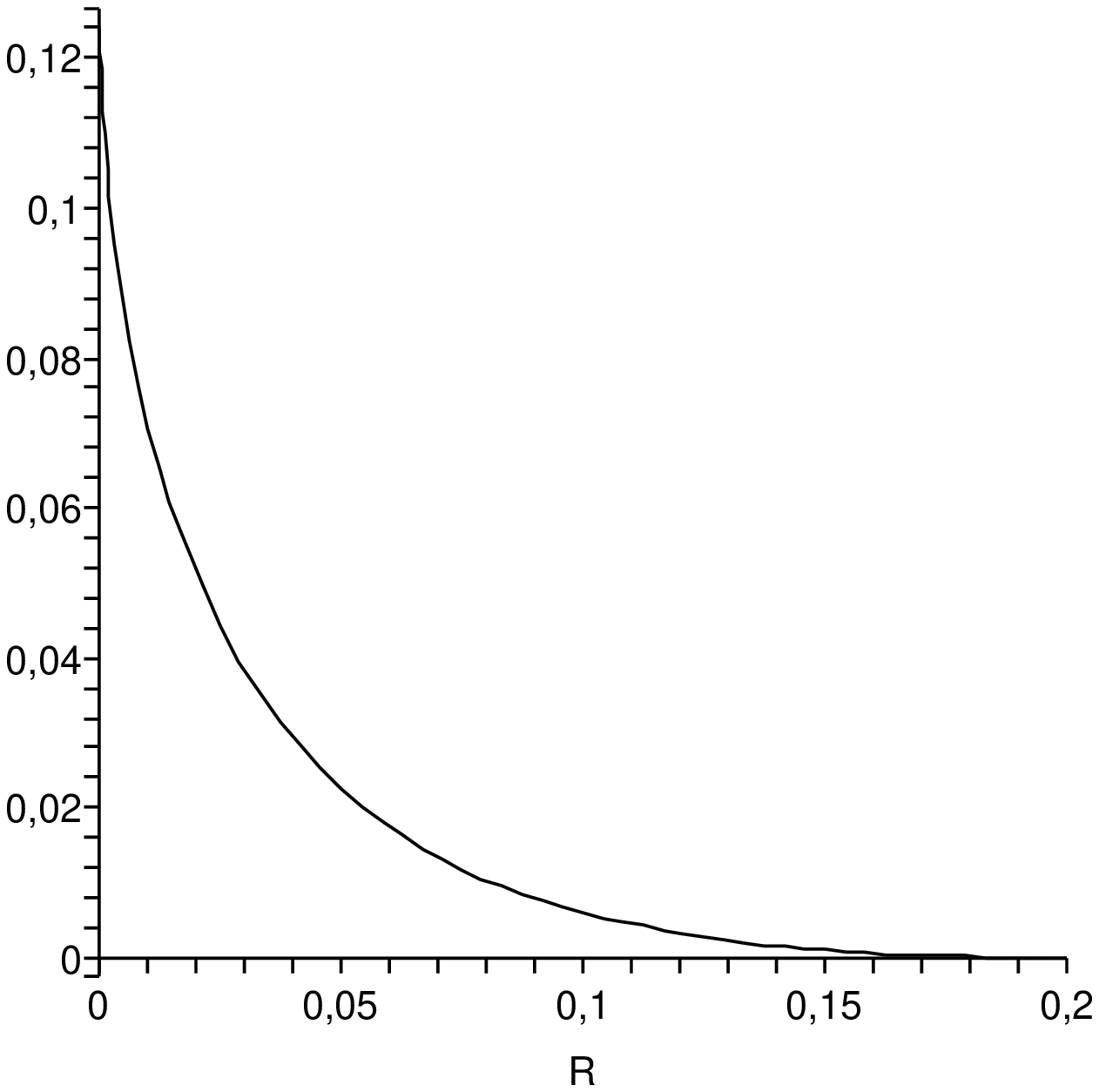}

\begin{center}
{Fig. 2. The plot of the function $p_{1}(R,0.01)$
($R_{\rm crit} \approx 0.387$).}
\end{center}

\medskip

\newpage

\includegraphics[width=0.9\hsize,height=0.8\hsize]{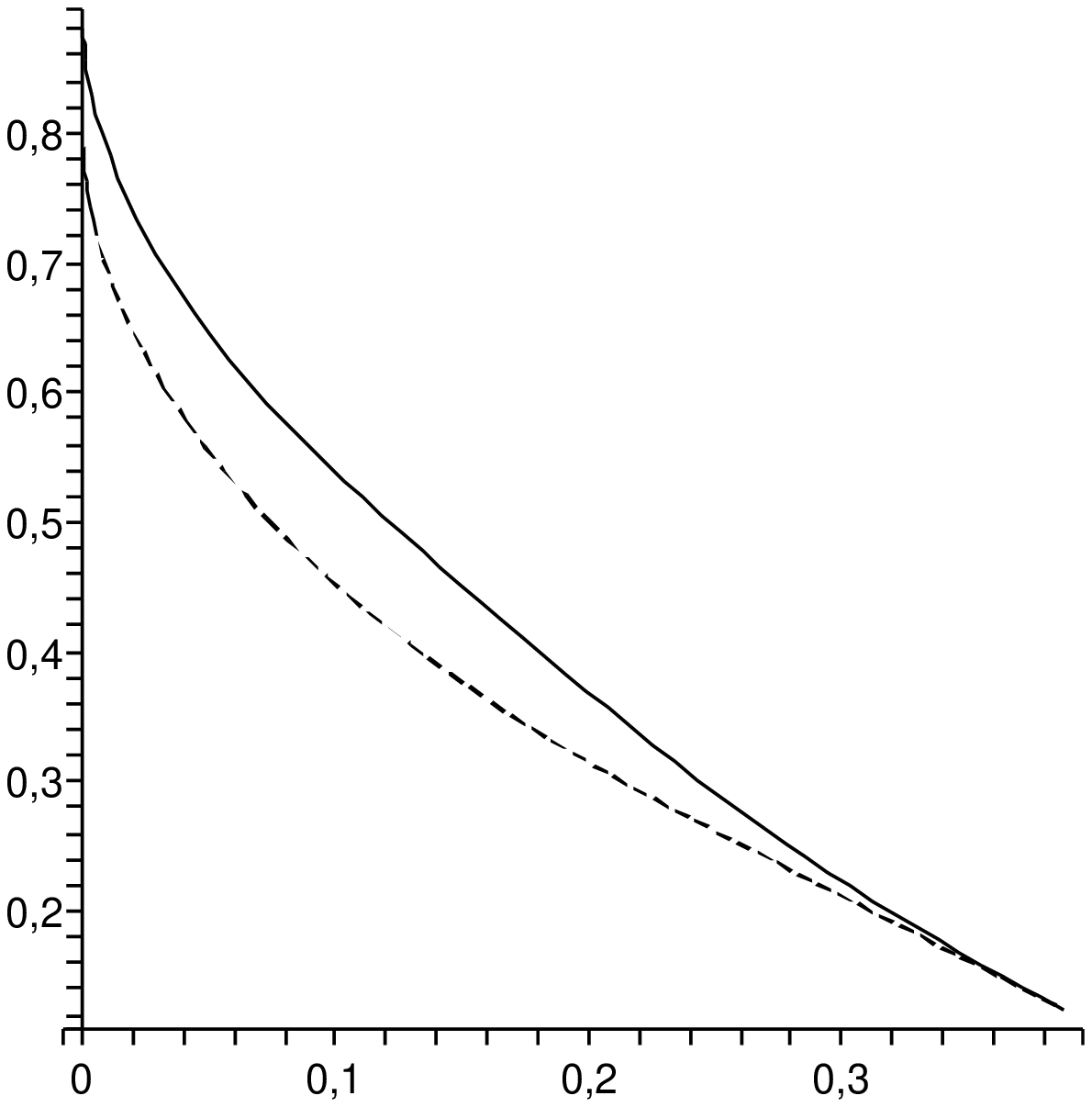}

\begin{center}
{Fig. 3. The plots of the functions $F_{1}(R,p)$ and
$E_{\rm ex}(R,p)$ for $p = 0.01$
($R_{\rm crit} \approx 0.387$).}
\end{center}

\newpage

\includegraphics[width=0.9\hsize,height=0.8\hsize]{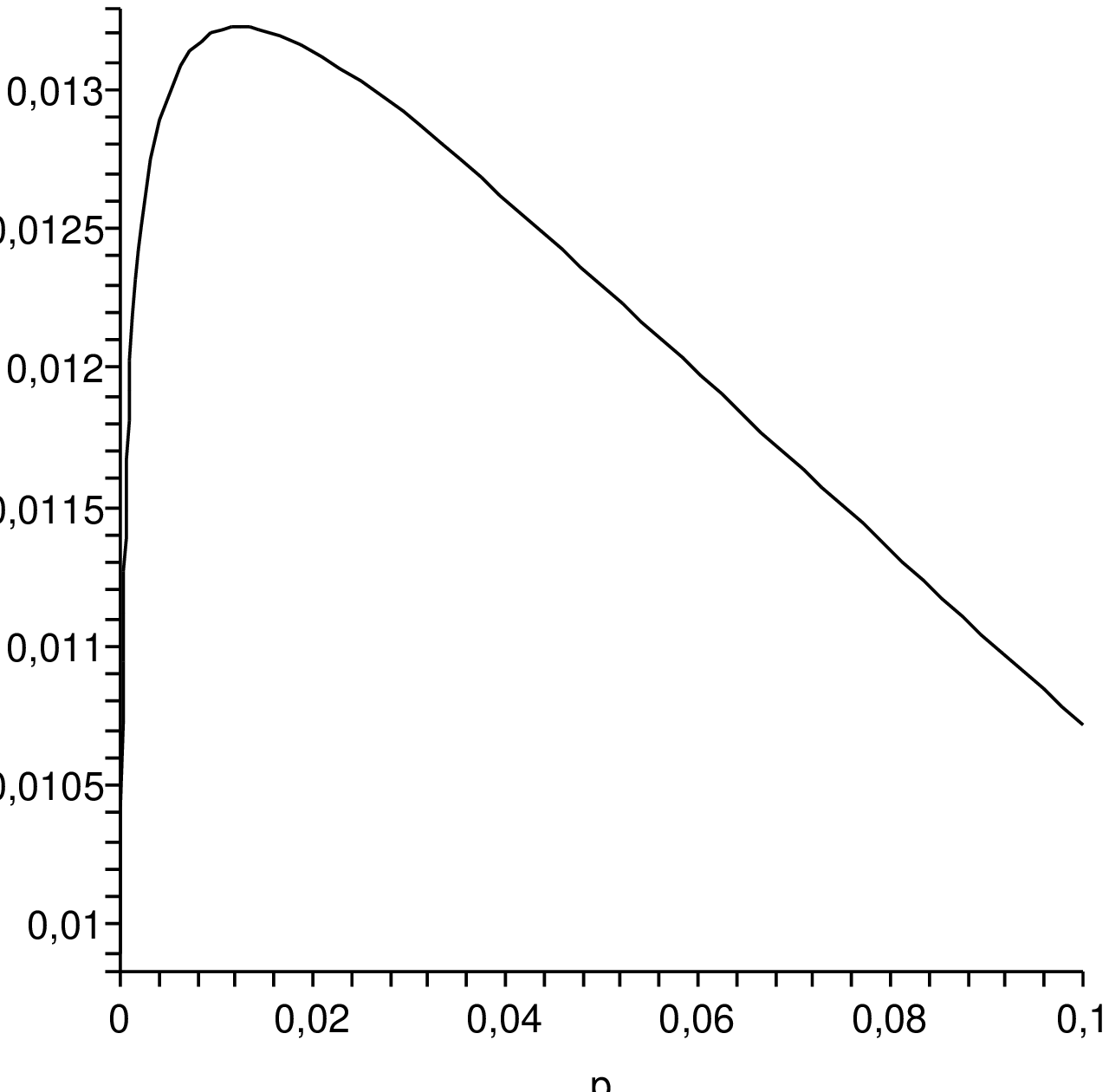}

\begin{center}
{Fig. 4. The plot of the function $p_{11}(p,0.1)$}
\end{center}

\end{document}